\documentclass[aps,pre,twocolumn,groupedaddress,showpacs]{revtex4-1}
\usepackage{amsmath,amssymb,color}
\usepackage{epsfig}
\begin{document}
\def\E{\mathbb{E}}
\def\P{\mathbb{P}}
\def\R{\mathbb{R}}
\def\Z{\mathbb{Z}}
\def\O{{\cal O}}
\def\scr{\scriptstyle}
\def\text{\textstyle}
\def\floor{\rm floor}
\def\sw{\rm sw}
\def\al{\alpha}
\def\be{\beta}
\def\ga{\gamma}
\def\del{\delta}
\def\la{\lambda}
\def\sig{\sigma}
\def\om{\omega}
\def\Om{\Omega}
\def\phi{\varphi}
\def\na{\nabla}
\def\A{{\cal A}}
\def\atanh{\rm atanh}
\def\tiz1{\tilde z_1}
\def\r{\mathbf{r}}
\def\u{\mathbf{u}}
\def\tirb{\tilde{\mathbf{r}}}
\def\ti{\tilde}
\def\p{\partial}
\def\sub{\subset}
\def\12{\frac{1}{2}}
\def\oraw{\overrightarrow}
\def\nea{\nearrow}
\def\sea{\searrow}
\def\beq{\begin{equation}}
\def\eeq{\end{equation}}
\def\beqn{\begin{equation*}}
\def\eeqn{\end{equation*}}
\def\beqa{\begin{eqnarray}}
\def\eeqa{\end{eqnarray}}
\def\beqan{\begin{eqnarray*}}
\def\eeqan{\end{eqnarray*}}
\title{Shape of pendant droplets under a tilted surface}
\author{Jo\"el De Coninck} \author{Juan Carlos Fern\'andez-Toledano} 
\affiliation{Laboratoire de Physique des Surfaces et Interfaces\\
Universit\'{e} de Mons, 20 Place du Parc, 7000 Mons, Belgium}
\author{Fran\c cois Dunlop} \author{Thierry Huillet} \author{Alvin Sodji}
\affiliation{Laboratoire de Physique Th\'{e}orique et Mod\'{e}lisation,
CNRS-UMR 8089\\ Universit\'{e} de Cergy-Pontoise, 
95302 Cergy-Pontoise, France}
\begin{abstract}
For a pendant drop whose contact line is a circle of radius $r_0$, we derive the
relation $mg\sin\al={\pi\over2}\ga r_0\,(\cos\theta^{\rm min}-\cos\theta^{\rm max})$
at first order in the Bond number, where $\theta^{\rm min}$ and $\theta^{\rm max}$
are the contact angles at the back (uphill) and at the front (downhill), $m$ is
the mass of the drop and $\ga$ the surface tension of the liquid.
The Bond (or E\"otv\"os) number is taken as
$Bo=mg/(2r_0\ga)$. The tilt angle $\al$ may increase from
$\al=0$ (sessile drop) to $\al=\pi/2$ (drop pinned on vertical wall) to
$\al=\pi$ (drop pendant from ceiling). The focus will be on pendant drops with
$\al=\pi/2$ and $\al=3\pi/4$. The drop profile is computed exactly, in the same
approximation. Results are compared with {\sl surface evolver} simulations,
showing good agreement up to about $Bo=1.2$, corresponding for example to
hemispherical water droplets of volume up to about $50\,\mu$L. 
An explicit formula for each contact angle $\theta^{\rm min}$ and $\theta^{\rm max}$
is also given and compared with the almost exact  {\sl surface evolver} values.
\end{abstract}
\maketitle
\section{Introduction}
The study of static contact angles and shapes of both sessile and pendant
drops on a substrate has been an important issue in engineering sciences,
including: drop condensation \cite{SBKM11}, biomedical (or
biological) microelectromechanical systems, drug delivery, to cite only a few.
Contact angles give information about wettability and surface energy. 
The three-phase contact angles of a liquid condensed on a
substrate is in direct relation with interfacial and body forces acting
on sessile or pendant drops. In spite of abundance of data on
contact angles for sessile drops, there is a gap in the knowledge of contact
angles of pendant drops on inclined surfaces. Pendant drops appear to be
difficult to deal with in experiments.

Drops on vertical window panes is an important issue to design self-cleaning
surfaces. Also the problem of a bubble pinned on top of a plate is amenable to
the one of a pendant drop, pinned underneath \cite{NADJ11,AQ02}.
Another application is pendant drop tensiometry where a droplet is suspended
from a needle \cite{BNDCT15}. 
Cheng et al. \cite{CLBRN90} performed experiments to measure
contact angles of pendant axisymmetric glycerin and water drops. When the
surface on which the pendant drop is deposited is inclined with respect to
the horizontal, the angle downhill is greater than the angle uphill.
These angles are a function of plate inclination. Overall, the drop is
deformed and becomes non-axisymmetric. Several authors \cite{BOS,EJ04I,DDH17}
have discussed the effect of plate inclination
on the contact angles of sessile drops. A sessile drop may be
understood as one in equilibrium resting on a flat surface while a pendant
drop is one which hangs from a ceiling or a wall. Though the equations
describing the
shape of both drops are similar, the body forces have opposing effects
tending in the former case to flatten the drop and in the latter one to
elongate its shape, possibly resulting in a neck region. The effect of plate
inclination breaks the axisymmetry of the problem. 

The effect of surface chemistry of solid-liquid combinations on contact
angles clearly is an active subject of research. Factors such as surface
energy, wettability and substrate vibration or oscillation \cite{MD18} also
affect contact angles. Temperature is also an independent control variable. 

Understanding the shape of a drop pinned on an incline has a long history in
Physics, starting with \cite{MO,Fr}.
An empirical relation between slope angle and contact angles was given by
Furmidge \cite{Fu}, and further studied by many authors, see \cite{EJ06,DDH17}
and references therein. 
More recently the case of pendant drops has attracted much interest
\cite{MS88,BKM12,MWI15}, \cite{MGKLSTV19} and references therein. 

In the present work, we limit ourselves to the analysis of the effect of
inclination and drop volume on contact angles for pinned pendant drops, which
are subject to the Laplace-Young equation. This is a non-linear partial
differential equation of second order, for a two-dimensional surface. More can
be done mathematically and numerically in the case of cylindrical symmetry,
$\al=0$ or $\pi$ with circular contact line \cite{OB91,FH13,BNDCT15}. There the
Laplace-Young equation takes the form of an ordinary differential equation. In
the present work we focus on the asymmetrical case, $\al\notin\{0,\pi\}$,
requiring partial differential equations techniques, illustrated with
$\al=\pi/2$ and $\al=3\pi/4$.
We perform the calculations for gravity but the technique can easily be adapted
to other distorting forces, associated with electric or magnetics fields.

In \cite{DDH17} we derived a linear response solution to the Laplace-Young
equation for a drop sitting on an incline, using cylindrical coordinates. In the
present paper we use spherical coordinates, covering a wider range of contact
angles, and including the pendant drop problem.

In Section \ref{LYsph} we give the setting in spherical cordinates and introduce
a linear response ansatz to solve the Laplace-Young equation at first order in
the Bond number. 
In Section \ref{furmidegerel} we show that the ansatz implies the Furmidge
relation described in the abstract and we test its validity against
{\sl surface evolver} simulations. A comparison is given between the pendant
drop ($\al=135\,$degrees) and the sessile drop ($\al=45\,$degrees).
In Section \ref{solution} we derive exact solutions for drop profiles at first
order in the Bond number and compare them with {\sl surface evolver}
simulations. In Section \ref{contact} we compute separately, in the same
approximation, each contact angle and compare the predictions with {\sl surface
evolver} results.
In Section \ref{evolver} we comment on the technical aspects of the approximate
numerical solution with {\sl surface evolver}.
In Section \ref{conclusion} we summarize the results and give a perspective.

%%%%%%%%%%%%%%%%%%%%%%%%%%%%%%%%%%%%%%%%%%%%%%%%%%%%%%%%%%%%%%%%%%%%%%%%%%%%%%%%
\section{Laplace-Young equation in spherical coordinates}\label{LYsph}
The Laplace-Young equation equates the Laplace pressure to the hydrostatic
pressure: 
\beq\label{LY}
p_{\rm liq}=p_{\rm{atm}}-2\gamma H=p_{0}+\rho \,{\bf g\cdot r}\,,\quad
H={1\over2}\left({1\over R_1}+{1\over R_2}\right)
\eeq
where $p_0$ is the pressure at the origin and $H$ is the mean curvature
defined with the outer normal pointing from the liquid into the atmosphere:
each principal radius of curvature is positive when the
corresponding center of curvature is on the side of the outer
normal and negative otherwise. 

The Cartesian frame of reference has $z$-axis perpendicular to the substrate,
pointing from the substrate into the liquid,
$x$-axis along the slope downwards, and $y$-axis horizontal, so that the
corresponding unit vectors satisfy ${\bf e_x\wedge e_y=e_z}$. The slope angle
$\alpha\in[0,\pi]$ corresponds to a rotation of angle $\al$ of the system and
frame of reference around the $y$-axis, so that the gravity vector
${\bf g}=(g\sin\al,\,0,\,-g\cos\al)$. Thus $\al=0$ is a sessile drop on a
horizontal substrate, $\al=\pi/2$ is a drop pinned on a vertical wall, $\al=\pi$
is a drop pendant from the ceiling. We use spherical polar coordinates with
origin at the center of the spherical cap at zero gravity,
%for $B=0$, %see Fig. 4 in \cite {DDH17}, 
$\theta \in [0,\theta _{0}]$ measured from the $z$-axis and azimuth
$\varphi \in [-\pi ,\pi ]$ measured from the $x$-axis. Then (\ref{LY}) reads
\beq\label{LY2}
p_{\rm{atm}}-2\ga H=p_{0}-\rho gr\cos\alpha\cos\theta
+\rho gr\sin\alpha\sin\theta\cos\varphi
\eeq
a partial differential equation for the drop profile $r(\theta,\varphi)$.
At zero gravity we have a spherical cap of volume $V$, radius
$R$ and contact angle $\theta_0$ such that $r_0=R\sin\theta_0$ and
\begin{align}
V&=\pi R^3\Bigl({2\over3}-\cos\theta_0+{1\over3}\cos^3\theta_0\Bigr)\cr
&={\pi R^3\over3}(1-\cos\theta_0)^2(2+\cos\theta_0)
\end{align}
The boundary condition for (\ref{LY2}) is
$r(\theta_0,\varphi)=R\ \forall\varphi$, and volume conservation implies
\beq
\int_0^{\theta_0}\sin\theta d\theta\int_{-\pi}^\pi d\varphi
\int_0^{r(\theta,\varphi)}r^2dr=V
\eeq
In order to convert (\ref{LY2}) to a dimensionless equation and reduce the
number of parameters, let
\beq
r=R\ti r\,,\qquad H=R^{-1}\ti H\,,\qquad p=\ga R^{-1}\ti p
\eeq
Then (\ref{LY2}) takes the form of the adimensional Laplace-Young equation
\beq\label{LY3}
\ti p_{\rm atm}-2\ti H=\ti p_{0}-B\ti r\cos\alpha\cos\theta
+B\ti r\sin\alpha\sin\theta\cos\varphi
\eeq
where $B$ is the Bond number associated to the length $R$,
\beq
B={\rho gR^2\over\ga}
\eeq
When comparing with {\sl surface evolver} simulations, we will use
\beq\label{Bo}
Bo={mg\over w\ga}={mg\over 2r_0\ga}
={B\pi(1-\cos\theta_0)^2(2+\cos\theta_0)\over6\sin\theta_0}
\eeq
where $w$ is the width of the drop basis, here equal to $2r_0$.
We have three independent parameters: $\al,\,\theta_0$, and $B$ or $Bo$. The
pressure difference $\ti p_0-\ti p_{\rm atm}$ will be determined from volume
conservation.

At small $B$, an approximate solution to (\ref{LY3}) may be obtained by a linear
response argument: the deformation from the spherical cap is caused by the two
terms linear in $B$. As functions of the azimuth $\phi$ these terms  generate a
two-dimensional vector space, linear combinations $a+b\cos\phi$ with $a,b$
functions of $\theta$. The linear response ansatz consists in looking for a
solution to (\ref{LY3}) in this vector space, at first order in $B$:
\begin{multline}\label{ansatz}
\ti r(\theta,\varphi)=1+B\,r_{01}(\theta)\cos\alpha
+B\,r_{11}(\theta )\sin\alpha\cos\varphi\cr+O(B^{2})  
\end{multline}
where $r_{01}(\theta)$ and $r_{11}(\theta)$ are arbitrary functions of $\theta$,
subject to the boundary conditions, and also subject to volume conservation at
first order in $B$.

In \cite{DDH17}, linear response was used in cylindrical coordinates, with
\begin{multline}\label{ansatzcyl}
z(r,\varphi)=z_{00}(r)+B\,z_{01}(r)\cos\alpha
+B\,z_{11}(r)\sin\alpha\cos\varphi\cr+O(B^{2})  
\end{multline}
where $z_{00}(r)$ is the spherical cap profile. The change of variables between
spherical and cylindrical coordinates is non-linear. Therefore the drop profiles
derived from linear response in either coordinates will differ by an $O(B^{2})$.

Also the drop volume is a linear functional of $z(r,\varphi)$ in cylindrical
coordinates, but a non-linear functional of $r(\theta,\varphi)$ in spherical
coordinates. This will also contribute a difference $O(B^{2})$ when imposing
volume conservation at first order in $B$.
%%%%%%%%%%%%%%%%%%%%%%%%%%%%%%%%%%%%%%%%%%%%%%%%%%%%%%%%%%%%%%%%%%%%%%%%%%%%%%%%
\section{The Furmidge relation}\label{furmidegerel}
%%%%%%%%%%%%%%%%%%
\begin{figure*}[tbp]
\begin{center}
\resizebox{16cm}{!}{\includegraphics{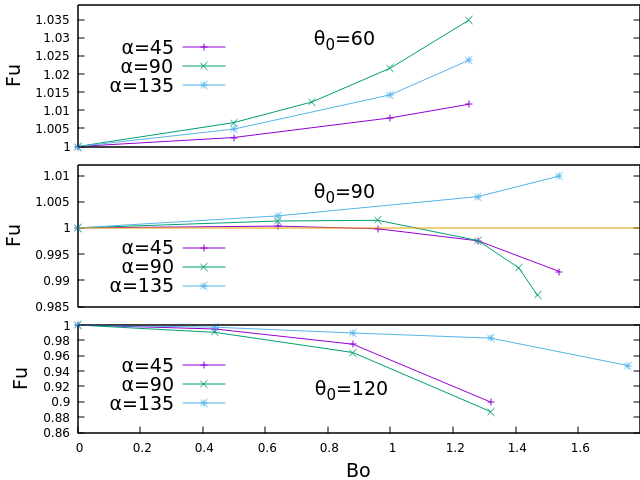}}
\end{center}
\caption{Furmidge ratio $Fu$ (\ref{furmidge}) for
$\theta_0=60,\,90,\,120\,$degrees
and $\al=45,\,90,\,135\,$degrees versus Bond number $Bo$ (\ref{Bo}).}
\label{t69}
\end{figure*}
%%%%%%%%%%%%%%%%%%
The ansatz (\ref{ansatz}) already implies the Furmidge relation \cite{Fu} as
stated in the abstract, following the argument in \cite{DDH17}.
Indeed let $\theta_\al(\phi)$ be the contact angle at azimuth $\phi$.
Then $\cos\theta_\al(\phi)$ may be computed from (\ref{ansatz}), using 
$r_{01}(\theta)$ and $r_{11}(\theta)$ and their derivatives at $\theta=\theta_0$.
The result as function of $\phi$, at first order in $B$, will again be in the
linear span of constants and $\cos\phi$. The values at $\phi=0$ and $\phi=\pi$
determine the linear combination:
\beq
\cos\theta_\al(\phi)={\cos\theta_\al^{\rm min}+\cos\theta_\al^{\rm max}\over2}
-\cos\phi{\cos\theta_\al^{\rm min}-\cos\theta_\al^{\rm max}\over2}%\cr+O(B^{2})
\eeq
where $\theta_\al^{\rm min}=\theta_\al(\pi)$ and $\theta_\al^{\rm max}=\theta_\al(0)$
are the contact angles at the back (uphill) and at the front (downhill).
This allows to compute the capillary force upon the drop, component parallel to
the substrate plane, a force which lies in the negative direction of
the $x$-axis:
\begin{align}
F_\ga&=\int_{-\pi}^{\pi}r_0d\phi\cos\phi\cos\theta_\al(\phi)\cr
&=-{\pi\over2}\ga r_0\bigl(\cos\theta_\al^{\rm min}-\cos\theta_\al^{\rm max}\bigr)
+O(B^2)
\end{align}
Balance with gravity implies the desired Furmidge relation,
\beq
mg\sin\alpha=
{\pi\over2}\ga r_0\bigl(\cos\theta_\al^{\rm min}-\cos\theta_\al^{\rm max}\bigr)
+O(B^2)
\eeq
We define the Furmidge ratio as
\beq\label{furmidge}
 Fu={\pi\over2}{\ga r_0\bigl(\cos\theta_\al^{\rm min}-\cos\theta_\al^{\rm max}\bigr)
  \over mg\sin\al}
    \eeq
obeying $Fu=1+O(B)$.

Therefore (\ref{furmidge}) with $Fu=1$ gives a prediction at first order in $B$
of $\cos\theta_\al^{\rm min}-\cos\theta_\al^{\rm max}$. We have simulated water
droplets with {\sl surface evolver}, and measured
$\theta_\al^{\rm min}$ and $\theta_\al^{\rm max}$, whence $Fu$,
plotted in Fig.~\ref{t69}.

The drop will de-pin when $\theta_\al^{\rm max}$ will increase up to the advancing
angle $\theta^{\rm advancing}$ or when $\theta_\al^{\rm min}$ will decrease down to
the receding angle $\theta^{\rm receding}$. Therefore the larger $Fu$ the closer to
de-pinning. $Fu>1$ means that the correction to linear response brings the drop
closer to the depinning instability.
This is also clear on the drop profiles, as in Fig. \ref{bx}.

%%%%%%%%%%%%%%%%%%%%%%%%%%%%%%%%%%%%%%%%%%%%%%%%%%%%%%%%%%%%%%%%%%%%%%%%%%%%%%%%
\section{Solution at first order in $B$.}\label{solution}
%%%%%%%%%%%%%%%%%%
\begin{figure*}[tbp]
\begin{center}
\resizebox{16cm}{!}{\includegraphics{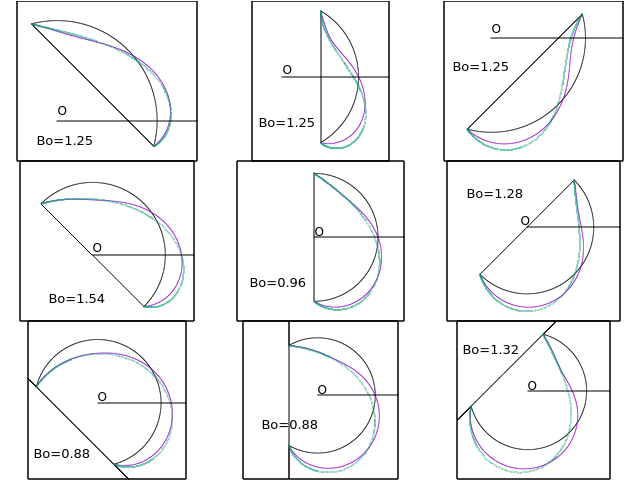}}
\end{center}
\caption{Droplet profiles for $\theta_0=60,\,90,\,120\,$degrees (from top to
  bottom) and $\al=45,\,90,\,135\,$degrees (from left to right), Bond numbers as
  shown. Green-blue points: surface evolver. Purple lines: linear response from
  (9)(27)(28). Black: spherical cap of same volume, centred at $O$, with contact
  angle $\theta_0$.}
\label{bx}
\end{figure*}
%%%%%%%%%%%%%%%%%%%%%%%%%%%%%%%%%%%%%%%%%%%%%%%%%%%%%%%%%%%%%%%%%%%%%%%%%%%%%%%%

The mean curvature $\ti H$ is given by
\beq
-2\ti H=\na\cdot{\bf n}
\eeq
The normal to a surface of equation $f(r,\theta,\varphi)=0$ is
\beq
{\bf n}=\na f\,/\,||\na f||
\eeq
Hence, using the gradient in spherical coordinates, and now denoting simply $r$
the adimensional radial coordinate,
\begin{multline}
{\bf n} =\nabla\,\Bigl[r-1-B\,r_{01}(\theta )\cos\alpha
-B\,r_{11}(\theta )\sin \alpha\cos\varphi\Bigr]\\
+O(B^{2}) \\
={\bf u}_{r}-B\,\Bigl({\frac{r_{01}^{\prime }(\theta )}{r}}\cos
\alpha +{\frac{r_{11}^{\prime }(\theta )}{r}}\sin \alpha \cos
\varphi \Bigr)\,{\bf u}_{\theta }\cr
+B\,{\frac{r_{11}(\theta )}{r\sin \theta}}\sin\alpha\sin\varphi\,{\bf u}_{\varphi }
+O(B^{2})
\end{multline}
where $||\na f||=1+O(B^2)$ has been used. Then
\begin{multline}
-2\ti H=\nabla \cdot {\bf n} \\
={\frac{2}{r}}-B\Bigl[{\frac{r_{01}^{\prime\prime }(\theta )}{r^{2}}}\cos\alpha
  +{\frac{r_{11}^{\prime \prime }(\theta )}{r^{2}}}\sin\alpha\cos\varphi\cr
  +{\frac{r_{01}^{\prime }(\theta )\cot \theta }{r^{2}}}\cos \alpha \\
+{\frac{r_{11}^{\prime }(\theta )\cot \theta }{r^{2}}}\sin \alpha
\cos \varphi -{\frac{r_{11}(\theta )}{r^{2}\sin ^{2}\theta }}\sin
\alpha \cos \varphi\Bigr]+O(B^{2}) \\
=-2\ti H_{0}-2B\ti H_{01}\cos\alpha -2B\ti H_{11}\sin \alpha \cos \varphi +O(B^{2})
\end{multline}
with, at $r=\tilde r(\theta,\varphi)$ as (\ref{ansatz}),
\begin{align}
-2\ti H_{0} &=2 \cr
-2\ti H_{01} &=-2r_{01}(\theta )-r_{01}^{\prime \prime }(\theta
)-r_{01}^{\prime }(\theta )\cot \theta \cr
-2\ti H_{11} &=-2r_{11}(\theta )-r_{11}^{\prime \prime }(\theta
)-r_{11}^{\prime }(\theta )\cot \theta +{\frac{r_{11}(\theta )}{\sin^{2}\theta }}\cr
\end{align}
Similarly 
\begin{equation}
\ti p_{0}=\ti p_{00}+B\ti p_{1}\cos \alpha +O(B^{2})
\end{equation}
Order 0 in $B$ for (\ref{LY3}) is the Laplace equation 
\begin{equation}
\ti p_{\rm{atm}}+2=\ti p_{00}
\end{equation}
Order one in $B$ for (\ref{LY3}) has terms independent of $\varphi $ and
terms linear in $\cos \varphi $, hence two ordinary differential equations 
\begin{eqnarray*}
-2\ti H_{01} &=&\ti p_{1}-\cos \theta \\
-2\ti H_{11} &=&\sin \theta
\end{eqnarray*}
or 
\begin{equation}
\begin{array}{l}
r_{01}^{\prime \prime }(\theta )+r_{01}^{\prime }(\theta )\cot \theta
+2r_{01}(\theta )=\cos \theta -\ti p_1 \\ 
r_{11}^{\prime \prime }(\theta )+r_{11}^{\prime }(\theta )\cot \theta
+r_{11}(\theta )(1-\cot ^{2}\theta )=-\sin \theta
\end{array}
\label{ode}
\end{equation}
for $\theta \in [0,\theta _{0}]$ with boundary conditions 
\begin{equation}
\begin{array}{l}
r_{01}^{\prime }(0)=0,\text{ }r_{01}(\theta _{0})=0 \\ 
r_{11}(0)=0,\text{ }r_{11}(\theta _{0})=0
\end{array}
\label{bc}
\end{equation}
Hence two parameters: $\theta _{0}\in [0,\pi ]$ and $\ti p_1\in \R $. The
ordinary differential equations (\ref{ode}) with boundary conditions (\ref
{bc}) are solved exactly with \textsl{Mathematica}: 
\begin{multline}
r_{01}(\theta)=\cos \theta [C_{01}-\frac{1}{3} \log (1+\cos \theta )]+ 
\frac{2-3\ti p_1}{6} \\
r_{11}(\theta ) =\sin \theta \,[C_{11}+{\frac{1}{3}}\log (1+\cos \theta )]+%
{\frac{\cos \theta -\cos ^{2}\theta }{3\,\sin \theta }}
\end{multline}
with $C_{11}$ from $r_{11}(\theta_{0})=0$,
\beq
C_{11}=-{\cos\theta_0\over3(1+\cos\theta_0)}-{1\over3}\log(1+\cos\theta_0)
\eeq
and $C_{01},\,\ti p_1$ from $r_{01}(\theta _{0})=0$ and volume conservation at
first order in $B$,
\beq
\int_0^{\theta_0}r_{01}(\theta) \sin\theta\,d\theta=0
\eeq
entailing
\beq
C_{01}={1\over6}+{1\over3}\log(1+\cos\theta_0)\ ;\
\ti p_1={2+\cos\theta_0\over3}
\eeq
Then
\begin{align}
  r_{01}(\theta)&={\cos\theta-\cos\theta_0\over 6}
  +\frac{\cos\theta}{3} \log{1+\cos\theta_0\over 1+\cos\theta}\label{r01} \\
  r_{11}(\theta )& ={\sin\theta\over 3}\left[{\cos\theta\over 1+\cos\theta}
    -{\cos\theta_0\over 1+\cos\theta_0}+\log{1+\cos\theta\over 1+\cos\theta_0}
    \right]\label{r11}
\end{align}
We can now easily plot the linear response drop profiles by inserting
(\ref{r01})(\ref{r11}) into (\ref{ansatz}) without $O(B^2)$, and compare with
{\sl surface evolver} profiles: Fig. \ref{bx},
computed for water droplets, $\rho=997\,$kg/m$^3$, $g=9.8\,$m/s$^2$,
$\ga=0.073\,$N/m.
The values of $Bo$ were chosen so that there is a significant difference between
the two. At half these Bond numbers the profiles almost coincide. At twice these
Bond numbers, the Laplace-Young equation under the given conditions does not
have a well defined solution.

\section{Contact angles}\label{contact}
%%%%%%%%%%%%%%%%%%
\begin{figure*}[tbp]
\begin{center}
\resizebox{16cm}{!}{\includegraphics{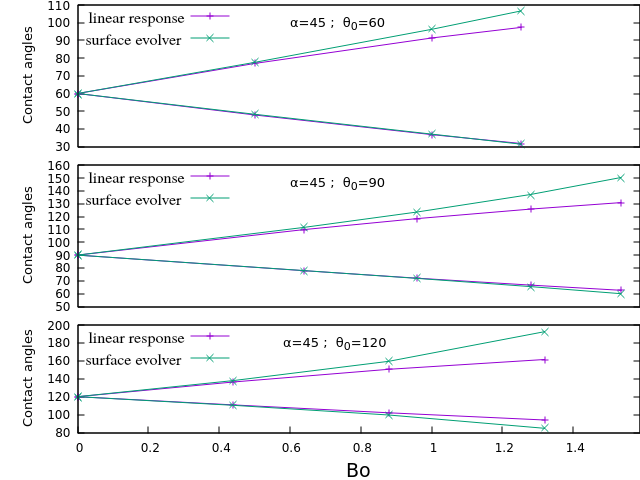}}
\end{center}
\caption{Contact angles $\theta^{\min}$ and $\theta^{\max}$ as function of $Bo$
  for $\al=45\,$degrees and $\theta_0=60,\,90,\,120\,$degrees. Linear response
  from (\ref{costheta}).}
\label{Fa45}
\end{figure*}
\begin{figure*}[tbp]
\begin{center}
\resizebox{16cm}{!}{\includegraphics{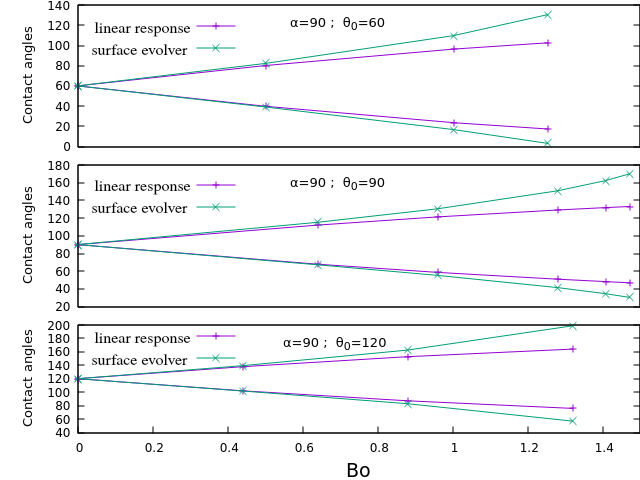}}
\end{center}
\caption{Contact angles $\theta^{\min}$ and $\theta^{\max}$ as function of $Bo$
  for $\al=90\,$degrees and $\theta_0=60,\,90,\,120\,$degrees. Linear response
  from (\ref{costheta}).}
\label{Fa90}
\end{figure*}
\begin{figure*}[tbp]
\begin{center}
\resizebox{16cm}{!}{\includegraphics{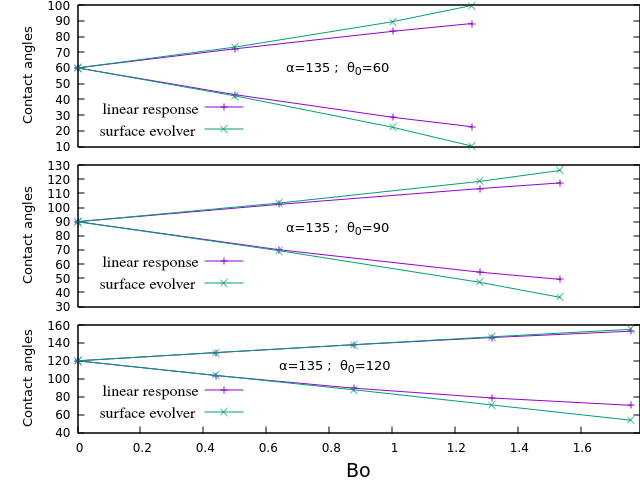}}
\end{center}
\caption{Contact angles $\theta^{\min}$ and $\theta^{\max}$ as function of $Bo$
  for $\al=135\,$degrees and $\theta_0=60,\,90,\,120\,$degrees. Linear response
  from (\ref{costheta}).}
\label{Fa135}
\end{figure*}

%%%%%%%%%%%%%%%%%%%%%%%%%%%%%%%%%%%%%%%%%%%%%%%%%%%%%%%%%%%%%%%%%%%%%%%%%%%%%%%%
The contact angle $\theta=\theta(\varphi )$ is given by 
\begin{equation}
  \cos\theta=\,\mathbf{u}_{\rho }\cdot
{\p\tilde{\mathbf{r}}/\p\theta\over|\p\tilde{\mathbf{r}}/\p\theta|}%
\bigg|_{\theta=\theta_{0}}
\end{equation}
where $\,\mathbf{u}_{\rho }=\cos\theta\,\mathbf{u}_{\theta}+\sin \theta\,
\mathbf{u}_{r}$ is the radial unit vector of plane polar coordinates
associated with the contact line circle.
We shall compute contact angles exactly for the linear response profile, namely
(\ref{ansatz}) without the $O(B^2)$ correction:
\beq
\tirb=\ti r(\theta,\phi)\u_r
\eeq
\beq
    {\p\tirb\over\p\theta}\bigg|_{\theta=\theta_0}=\u_\theta+
B(r_{01}^{\prime }(\theta_0)\cos\alpha+r_{11}^{\prime }(\theta_0)\sin\alpha\cos\phi)
    \u_r
\eeq
\beq\label{costheta}
\cos\theta={\cos\theta_0+
  B(r_{01}^{\prime }(\theta_0)\cos\alpha+r_{11}^{\prime }(\theta_0)\sin\alpha\cos\phi)
  \sin\theta_0\over(1+B^2(r_{01}^{\prime }(\theta_0)\cos\alpha
  +r_{11}^{\prime }(\theta_0)\sin\alpha\cos\phi)^2)^{1/2}}
\eeq
with, from (\ref{r01})(\ref{r11}),
\[
r_{01}^{\prime }(\theta_0 )=-\sin \theta_0 /6+\sin \theta_0 \cos\theta_0 /(3(1+\cos
\theta_0 )) 
\]
\[
r_{11}^{\prime }(\theta_0 )=\cos\theta_0 /3-2/(3(1+\cos\theta_0 )) 
\]
Then $\cos\theta^{\max}$ is given by (\ref{costheta}) with $\phi=0$
and $\cos\theta^{\min}$ with $\phi=\pi$. The corresponding contact angles
$\theta^{\min}$ and $\theta^{\max}$ are plotted for $\alpha=45\,$degrees on
Fig.~\ref{Fa45},  for $\alpha=90\,$degrees on Fig.~\ref{Fa90},  for
$\alpha=135\,$degrees on Fig.~\ref{Fa135}, together
with the almost exact {\sl surface evolver} values.
Increasing $Bo$ at constant $\theta_0$ corresponds to scaling up all length
scales, drop and its basis alike.

The dimensionless variables $Bo$ (ratio of gravity over capillarity)
and $\theta_0$ (contact angle at zero gravity) may be related to $V$ and $r_0$
through
\begin{align}
  Bo&={\rho gV\over2\ga r_0}\cr
{(1-\cos\theta_0)^2(2+\cos\theta_0)\over\sin^3\theta_0}&={3V\over\pi r_0^3}
\end{align}
or
\begin{align}
  r_0^2&=
 {6\ga\over\pi\rho g}{Bo\sin^3\theta_0\over(1-\cos\theta_0)^2(2+\cos\theta_0)}\cr
  V&={2\ga r_0\over\rho g}Bo
\end{align}
%%%%%%%%%%%%%%%%%%
%%%%%%%%%%%%%%%%%%%%%%%%%%%%%%%%%%%%%%%%%%%%%%%%%%%%%%%%%%%%%%%%%%%%%%%%%%%%%%%%
%%%%%%%%%%%%%%%%%%%%%%%%%%%%%%%%%%%%%%%%%%%%%%%%%%%%%%%%%%%%%%%%%%%%%%%%%%%%%%%%
\section{\sl Surface evolver}\label{evolver}
  The approximate numerical solutions of the Laplace-Young equation under the
  given conditions were obtained with the finite elements software
  {\sl Surface evolver}. A fine mesh implies slow convergence in time (number of
  iterations) due in particular to the vicinity of zero modes (capillary waves),
  weakly damped
  by gravity. We used linear elements, for which convergence in the number of
  elements or vertices goes like the inverse of that number, again a slow
  convergence. The asymptotic values of the Furmidge coefficient $Fu$ were
  obtained by a least squares fit to $a+b/$(number of vertices), with up to
  82000 vertices. Altogether each point in Fig.~\ref{t69}, or each
  drop profile in Fig.~\ref{bx} required about 15h of CPU.
%%%%%%%%%%%%%%%%%%
%%%%%%%%%%%%%%%%%%%%%%%%%%%%%%%%%%%%%%%%%%%%%%%%%%%%%%%%%%%%%%%%%%%%%%%%%%%%%%%%
%%%%%%%%%%%%%%%%%%%%%%%%%%%%%%%%%%%%%%%%%%%%%%%%%%%%%%%%%%%%%%%%%%%%%%%%%%%%%ùùù
  \section{Conclusion}\label{conclusion}
We have studied pendant drops pinned under an incline of tilt angle $\alpha
\in \left[ \pi /2,\pi \right) $, with a circular contact line and contact
angle $\theta _{\alpha }(\phi )$ at azimuth $\phi $ obeying
\[0\le \theta ^{\rm{receding}}\le \theta _{\alpha }^{{\rm min}}\le \theta
_{\alpha }(\phi )\le \theta _{\alpha }^{{\rm max}}\le \theta ^{\rm{%
advancing}}\le \pi ,
\]
thus for a very large range of contact angles. We have illustrated the
results with $\alpha =\pi /2$ and $\alpha =3\pi /4.$

In such situations, the axisymmetry of the problem is broken and the shape
of the drops can only be handled by the non-linear Laplace-Young partial
differential equation. We developed a linear response ansatz leading to an
exact integrable solution of the Laplace-Young equation at first order in a
Bond number. The use of spherical coordinates was shown particularly
relevant for the pendant drop problem. Comparison of the obtained
approximate drop profiles with {\it surface evolver} simulations showed good
agreement for Bond numbers up to about $Bo=1.2$, corresponding for example to
hemispherical water droplets with volume up to about $50\,$mm$^{3}$. The linear
response ansatz has also been shown
to imply the (small Bond number) Furmidge-like relation described in the
abstract, translating a balance between capillary force and
gravity; the validity of this relation was successfully tested against {\it %
surface evolver} simulations. It would be interesting to better identify the
upper value of the Bond number at which this relation starts failing.
We also derived a small Bond number expression for the contact angles,
showing good agreement with the simulations.

Sessile drops are less sensitive to gravity than pendant drops. For
comparison with the earlier work \cite{DDH17} therefore, we included the study
of a sessile drop pinned on an incline of tilt angle $\alpha =\pi /4.$%
\bigskip
\begin{acknowledgments}
  The authors thank the European Space Agency (ESA) and the Belgian Federal Science Policy Office (BELSPO) for their support in the framework of the PRODEX Programme.
This research was also partially funded by the Inter-University Attraction Poles 
Programme (IAP 7/38 MicroMAST) of the Belgian Science Policy Office, FNRS and R\'egion Wallonne.
\end{acknowledgments}

\bibliographystyle{apsrev4-1}

%\bibliography{ddh19}
\end{document}